\documentclass[showpacs,amsmath,amssymb,twocolumn,pra]{revtex4-1}

\usepackage[dvips]{graphicx} 
\usepackage{amsfonts}
\usepackage{amssymb}
\usepackage{amscd}
\usepackage{amsmath}    
\usepackage{enumerate}
\usepackage{epsfig}
\usepackage{subfigure}
\usepackage{bm}
\usepackage{xcolor}
\usepackage{amsthm}
\usepackage{framed}
\usepackage{multirow}
\usepackage{mathrsfs,amssymb}
\usepackage{latexsym} 
\usepackage{framed}
\usepackage{physics}
\usepackage{epstopdf}
\usepackage[skins]{tcolorbox}
\newtcolorbox[auto counter]{tbox}[2][]{%
    enhanced, float=hbt, drop fuzzy shadow southeast,
    colback=white!5!white, colframe=white!50!black,
    width= .97\columnwidth,sharp corners, boxrule=0.8pt,
    title={Table \thetcbcounter: #2}, #1
}

\begin{document}
\title{Continuous-Variable Source-Independent Quantum Random Number Generator with a Single Phase-Insensitive Detector}

\begin{abstract}
Quantum random number generators (QRNGs) harness quantum mechanical unpredictability to produce true randomness, which is crucial for cryptography and secure communications. Among various QRNGs, source-independent QRNGs (SI-QRNGs) relax the trust on the quantum source, allowing for flexible use of advanced detectors to achieve high randomness generation rates. Continuous-variable (CV) SI-QRNGs, in particular, hold promise for practical deployment due to their simplicity and randomness generation rates comparable to trusted-device QRNGs. In this work, we propose a novel CV-SI-QRNG scheme with a single phase-insensitive detector, and provide security proof based on semi-definite programming (SDP). We introduce a dimension reduction technique, which rigorously reduces an infinite-dimensional SDP problem to a finite-dimensional one, enabling efficient computation while maintaining valid randomness lower bound. We further validate our method through simulations. These results demonstrate the feasibility of our framework, paving the way for practical and simple SI-QRNG implementations.
\end{abstract}

\author{Hongyi Zhou}
\email{zhouhongyi@ict.ac.cn}
\affiliation{State Key Lab of Processors, Institute of Computing Technology, Chinese Academy of Sciences, 100190, Beijing, China.}

\maketitle
\section{introduction}
Quantum random number generators (QRNGs) exploit the inherent unpredictability of quantum mechanics to generate true randomness, which is essential for cryptographic applications and secure communications \cite{RMPQRNGReview16,MaQRNGReview16}. Among the various QRNG designs, semi-device-independent (semi-DI) QRNGs strike a balance between practical implementation and high security by partially trusting the devices \cite{Banik15,cao2015loss,brask2017megahertz,Nie16,li2019QRNG,Ma16,vsupic2017measurement,bischof2017measurement,PhysRevLett.118.060503,PhysRevX.10.041048,PhysRevApplied.15.034034,avesani2018source,smith2019simple}. A notable subclass is source-independent (SI) QRNGs, where the randomness is independent of the trustworthiness of the quantum source \cite{Ma16,smith2019simple,PhysRevLett.118.060503,avesani2018source}. This allows for flexibility in the choice of measurement devices, i.e., employing advanced detectors to extract more than one bit of randomness per sample. Recent advances in continuous-variable (CV) SI-QRNG protocols \cite{smith2019simple,PhysRevLett.118.060503,avesani2018source} have achieved randomness generation rates up to Gbps, comparable to those of trusted-device QRNGs, making CV-SI-QRNGs highly promising for practical deployment.

In traditional SI-QRNG designs, security is based on the entropic uncertainty relation, which relies on conjugate projective measurements applied to the untrusted quantum source. These measurements, such as Pauli $X$ and $Z$ operators \cite{Ma16} or $Q$ and $P$ quadratures in phase space \cite{PhysRevLett.118.060503}, are fundamental to quantum mechanics. However, the limited choices of conjugate measurements constrain the design of SI-QRNG protocols, especially for scenarios requiring high-dimensional or continuous-variable states.

A more general design for SI-QRNG involves replacing conjugate projective measurements with general positive operator-valued measures (POVMs) \cite{avesani2018source,avesani2022unbounded}, which significantly expands the range of implementable protocols. This generalization is particularly well-suited for numerical methods in security analysis \cite{zhou2023numerical}. Numerical frameworks transform the problem of calculating the randomness lower bound into semi-definite programming (SDP) problems. Compared with analytical methods, these numerical approaches are more versatile, accommodating a broader range of QRNG designs while usually providing tighter security bounds.

Despite its advantages, the numerical security analysis of CV-SI-QRNGs faces a major challenge: the quantum signals are typically continuous-variable optical states, such as coherent states, which are infinite-dimensional in the Fock basis. Moreover, the POVMs used for their measurement are also infinite-dimensional, making direct numerical analysis infeasible due to the necessity of solving infinite-dimensional SDP problems. Existing approaches address this challenge using squashing models \cite{PhysRevLett.101.093601}, which map infinite-dimensional systems onto finite-dimensional virtual protocols. However, squashing models impose restrictive assumptions on the POVMs, excluding many practical implementations \cite{PhysRevLett.101.093601}. Generalized squashing models mitigate this limitation but often result in overly pessimistic randomness bounds \cite{gittsovich2014squashing}.

In this work, we propose a numerical framework for the security proof of CV-SI-QRNGs that avoids the squashing model entirely. By introducing a novel dimension-reduction technique \cite{upadhyaya2021dimension}, we rigorously bound the infinite-dimensional SDP with a finite-dimensional counterpart, enabling efficient computation while maintaining tight randomness bounds. Additionally, we identify a class of POVMs compatible with our framework, characterized by diagonal elements in the Fock basis. This class includes many phase-insensitive detectors, such as single-photon detectors, allowing for the realization of a CV-SI-QRNG using a single detector. This design greatly simplifies practical implementation. We validate our approach through simulations with a time-multiplexed single-photon detector, demonstrating its feasibility and potential for high-performance applications.

\section{Preliminaries}

\subsection{Semi-definite programming}
The main technique applied in this work is the numerical security analysis framework based on semi-definite programming (SDP), which is a powerful tool widely used in quantum information science. An SDP problem is typically expressed in the following standard form,
\begin{equation}
\begin{aligned}
& \max_{X} \mathrm{tr}(C X)  \\
 \text{s.t.} \quad 
 & \mathrm{tr}(A_i X) = b_i \; \forall i, \\
 & X \succeq 0 \\
\end{aligned}
\end{equation}
where $X$ is the optimization variable, $C$ and $A_i$ are given Hermitian matrices, and $b_i$ are real constants. The constraint $X\succeq 0$ indicates that $X$ is positive semi-definite. The SDP problem is a special type of convex optimization problem which can be efficiently solved in polynomial time.

\subsection{Numerical framework for SI-QRNG}
The numerical framework for source-independent quantum random number generators (SI-QRNGs) utilizes SDP to calculate the guessing probability under an adversary’s optimal strategy. The state of the unknown quantum source $\rho$ can be expressed as a pure-state decomposition
\begin{equation}
\rho = \sum_i q_i \ket{\psi_i}\bra{\psi_i}.
\end{equation}
The adversary's goal is to maximize the guessing probability, which is defined as
\begin{equation}
\max_{q_i, \ket{\psi_i}} p_{\mathrm{guess}} = \max_{q_i, \ket{\psi_i}} \sum_i q_i \max_j \mathrm{tr}(\ket{\psi_i}\bra{\psi_i}M_j).
\end{equation}
To simplify the computation, we group the pure states into $n$-groups such that a pure state in the $k$-th group satisfies $\max_j \mathrm{tr}(\ket{\psi_i}\bra{\psi_i}M_j) =\mathrm{tr}(\ket{\psi_i}\bra{\psi_i} M_{k})$. Then the guessing probability is simplified into
\begin{equation}
\max_{\rho_k} p_{\mathrm{guess}} = \max_{\rho_k} \sum_{k=1}^n \mathrm{tr}(\rho_k M_k),
\end{equation}
where $\rho_k$ is a sub-normalized quantum state $\rho_k =\sum_{i\in S_k}q_i\ket{\psi_i}\bra{\psi_i}$.
This allows us to reformulate the problem as an SDP with the following constraints
\begin{equation}\label{eq:SDPprimalSI}
\begin{aligned}
& \max_{\rho_k} \sum_{k=1}^n \mathrm{tr}\left(\rho_k M_k\right) \\
\mathrm{s.t.} \quad & \mathrm{tr}\left(M_j\sum_{j=1}^n\rho_j\right) = p_j \quad \forall j, \\
& \mathrm{tr}\left(\sum_{k=1}^n \rho_k\right) =1, \\
& \rho_k \succeq 0 \quad \forall k,
\end{aligned}
\end{equation}
where the first constraint means the unknown source should be compatible with the experimental statistics and the second constraint is the normalization condition.
By solving this SDP, we obtain an upper bound on the guessing probability $p^U_{\text{guess}}$. The randomness is quantified by the conditional min-entropy $H_{\mathrm{min}}(A|E) \geq -\log_2 p^U_{\mathrm{guess}}$.

The framework described above assumes that the Hilbert space dimension of the unknown quantum state matches that of the trusted POVMs, significantly simplifying the numerical analysis. However, this assumption may not hold in practical implementations, especially for continuous-variable (CV) systems where the states and measurements are inherently infinite-dimensional. Addressing this discrepancy is critical to ensure accurate randomness quantification for CV-SI-QRNGs.

\section{CV-SI-QRNG protocol with a phase-insensitive detector}
\begin{figure*}[hbt]
\centering
\includegraphics[width=0.9\linewidth]{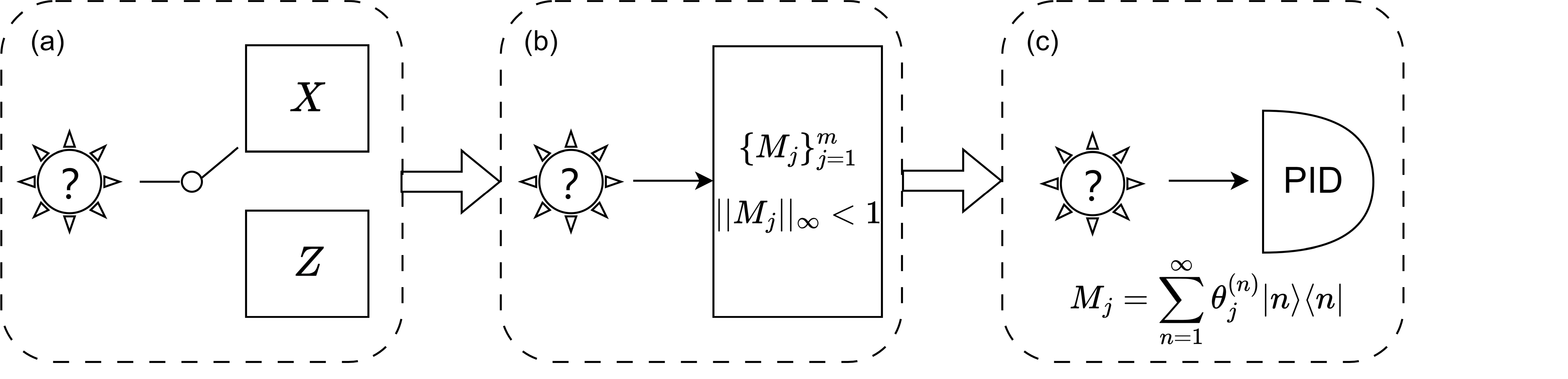}
\caption{(a) Conventional SI-QRNG protocols based on conjugate projective measurements. Here we take Pauli $X$ and $Z$ measurements as an example. (b) SI-QRNG protocol based on non-projective POVM, which is a generalization of conventional SI-QRNG protocols. (c) Our CV-SI-QRNG protocol is a specialization of SI-QRNG based on non-projective POVM, employing a single phase-insensitive detector. PID: phase-insensitive detector.}
\label{fig:protocol}
\end{figure*}

The basic idea of our protocol is illustrated in Fig.~\ref{fig:protocol}. The original SI-QRNG protocol is based on two conjugate projective measurements, for example, Pauli $X$ and Pauli $Z$ measurement. The user randomly switches between the two measurements. The randomness lower bound is given by entropic uncertainty relation. We notice that this randomly switched projective measurement form a POVM, i.e., $\frac{1}{2}\{\ket{0}\bra{0},\ket{1}\bra{1},\ket{+}\bra{+},\ket{-}\bra{-}\}$.
Actually for general POVM $\{M_j\}_{j=1}^m$, nonzero randomness can be generated for arbitrary unknown quantum state as long as the POVM is not a projective measurement, i.e., the spectrum norm of an arbitraty POVM element $M_j$ should be less than 1. This observation allows us to design more general SI-QRNG protocols. In optical experiments, phase-insensitive detectors, such as avalanche photodiodes, superconducting nanowire single-photon detectors, and other types of photo detectors,  
are widely used for obtaining photon number statistics or mean photon number. The absolute phase of the light will not affact the measurement outcomes.
The POVM element of a phase-insensitive detection $M_j$ is given by an infinite dimensional matrix in Fock-state basis,
\begin{equation}
M_j = \sum_{n=0}^\infty \theta_{j}^{(n)}\ket{n}\bra{n}.
\end{equation}
Then it satisfies a non-projective measurement as long as $\theta_{j}^{(n)}<1$.
In the following, we use phase-insensitive detectors to design a CV-SI-QRNG protocol.
The protocol is described as follows. 
\begin{enumerate}
	\item The untrusted source sends an unknown quantum state to a phase-insensitive detector.
	\item The detector outputs a measurement result $j\in\{1,2,\dotsc, m\}$, characterized by a set of POVM $\{M_j\}_{j=1}^m$. 
	\item After repeating steps 1-2 for a large number of rounds, Alice records the probability of each outcome $p_j$ and estimate the mean photon number $\langle n\rangle$. 
	\item Alice calculates the upper bound of the probability of successful guessing. Then she performs the post-processing to extract the final random numbers.
\end{enumerate}
The protocol description is general for all SI-QRNGs. For a CV-SI-QRNG, the key difference is that the dimension of $M_j$ is infinity, which make the security analysis difficult. One possible method to reduce the dimension is squashing model, which is to find a completely positive and trace-preserving map $\Lambda$, such that
\begin{equation}
\mathrm{tr}(\rho M_j) = \mathrm{tr}(\Lambda(\rho) M^\prime_j) \quad \forall j,
\end{equation}
holds for any input state $\rho$, where $\Lambda$ can project an infinite dimensional quantum state into a finite dimensional Hilbert space and $\{M^\prime_j\}_j$ is finite dimensional POVM. If a squashing model exists, we can substitute the squashed POVM $\{M^\prime_j\}_j$ into the SDP problem Eq.~\eqref{eq:SDPprimalSI} and the numerical security analysis becomes feasible.  

In this work, we will use another simpler approach, a novel dimension reduction technique based on the diagonal form of POVM of phase-insensitive detectors to deal with the infinite dimensional optimization problem.

\section{Security analysis}

Recalling Eq.~\eqref{eq:SDPprimalSI}, we can see that the security analysis of source-independent quantum random number generators (SI-QRNGs) based on general positive operator-valued measures (POVMs) is particularly well-suited for numerical approaches. The form of POVM and probabilities for each measurement output are exactly what we need to construct a SDP problem. The probabilities of each outcome, derived from the unknown quantum state, provide the necessary constraints. In the following, we will introduce how to reduce the dimension to make the computation feasible.

Notice that for our CV-SI-QRNG protocol, Eq.~\eqref{eq:SDPprimalSI} is an infinite dimensional SDP problem. To make it computable, we define a projection on 
the subspace where the photon number is less than $N$, i.e., $P = \sum_{n=0}^{N-1}\ket{n}\bra{n}$, and its complement $\bar{P} = I-P$. Then after the projection $P$, all operators in Eq.~\eqref{eq:SDPprimalSI} are represented by $N$-dimensional matrices in Fock basis.

We try to find an upper bound of the target function. Suppose the solution to Eq.~\eqref{eq:SDPprimalSI} is $\rho_k^*$, we have the following relation of the maximum guessing probability,
\begin{equation}\label{eq:upper_bound_finite}
\begin{aligned}
\sum_{k=1}^m \mathrm{tr}(\rho^*_k M_k) & = \sum_{k=1}^m \mathrm{tr}(\rho^*_{k,N}  M_{k,N}) + \sum_{k=1}^m \mathrm{tr}(\rho^*_{k,\bar{N}} M_{k,\bar{N}}) \\
& = \sum_{k=1}^m \mathrm{tr}(\rho^*_{k,N}  M_{k,N}) + 1 - \sum_{k=1}^m \mathrm{tr}(\rho^*_{k,N}) \\
& \leq \max_{\rho_{k,N} \in S_N} \sum_{k=1}^m \mathrm{tr}(\rho_{k,N}   M_{k,N}) + 1 - \sum_{k=1}^m \mathrm{tr}(\rho_{k,N}).
\end{aligned}
\end{equation}
where the subscript $N$ of an operator $O$ represents a projection $POP$ while the subscript $\bar{N}$ represents $\bar{P}O\bar{P}$. One needs to choose a suitable $S_N$ such that $P_N S_\infty P_N \subseteq S_N$, where $S_\infty$ is the feasible set of Eq.~\eqref{eq:SDPprimalSI}.


Next we construct the constraints for optimization problem after the projection to the photon number subspace. For the normalization condition, it is straightforward that $\mathrm{tr}\left(\sum_{k=1}^m \rho_{k, N}\right) \leq 1 $. For the probabilities of each measurement outcome, we have the following relations,
\begin{equation}\label{eq:holder}
    \begin{aligned}
       &\mathrm{tr}\left(M_j\sum_{k=1}^m\rho_k\right) \\
       =& \mathrm{tr}\left(M_{j,N}\sum_{k=1}^m\rho_{k,N}\right)+\mathrm{tr}\left(M_{j,\bar{N}}\sum_{k=1}^m\rho_{k,\bar{N}}\right) \\
       \leq&  \mathrm{tr}\left(M_{j,N}\sum_{k=1}^m\rho_{k,N}\right)+ ||\sum_{k=1}^m\rho_{k,\bar{N}} ||_1 || M_{j,\bar{N}}||_\infty.
    \end{aligned}
\end{equation}
In the first equality we apply the property that $[M_j,P] = 0$ since $M_j$ is diagonal in Fock state representation. In the second inequality
we consider Hölder’s inequality, i.e., $\mathrm{tr}{(A^\dag B)}\leq ||A||_p ||B||_q$ where $1/p+1/q =1$, and choose $p=1$ and $q=\infty$.
Notice that $||\sum_{k=1}^m\rho_{k,\bar{N}} ||_1$ is the weight outside the subspace with photon number less than $N$.
What we need is its upper bound. We consider the following optimization problem,
\begin{equation}\label{eq:weight_upper_bound}
    \begin{aligned}
       & \max_{\rho\in S_\infty} \mathrm{tr}(\rho P_{\bar{N}}) \\
       \mathrm{s.t.} \quad 
       &  \mathrm{tr}(\rho a^\dag a) = \langle n \rangle\\
       & \mathrm{tr}\rho =1 .
    \end{aligned}
\end{equation}
The first constraint represents the mean photon number measurement. We assume the mean photon number can be well estimated by a phase-insensitive detector. 
We consider the dual problem of Eq.~\eqref{eq:weight_upper_bound},
\begin{equation}\label{eq:weight_upper_bound_dual}
\begin{aligned}
    & \min_{ x, y \in \mathbb{R} } -x - y \langle n\rangle \\
    \mathrm{s.t.} \quad &  P_{\bar{N}} - xI - y a^\dag a \preceq 0.
\end{aligned}
\end{equation}
A feasible solution to Eq.~\eqref{eq:weight_upper_bound} is enough to ensure the security. We can quickly find a feasible solution that $x=0$ and $y=1/N$. Then an upper bound of the weight outside the subspace is $n/N$. 

Combining Eqs.~\eqref{eq:upper_bound_finite} and \eqref{eq:holder}, we are able to construct the optimization problem in the subspace with photon number less than $N$,
\begin{equation}\label{eq:sdp_pri_finite}
\begin{aligned}
&\max_{\rho_{k,N} \in S_N} \sum_{k=1}^m \mathrm{tr}(\rho_{k,N}   M_{k,N}) + 1 - \sum_{k=1}^m \mathrm{tr}(\rho_{k,N} ) \\
 \mathrm{s.t.} \quad &
    p_j - \frac{\langle n \rangle}{N} || M_{j,\bar{N}}||_\infty \leq \mathrm{tr}\left(M_{j,N}\sum_{k=1}^m\rho_{k,N}\right) \leq p_j  \\
    & \mathrm{tr}\left(\sum_{k=1}^m \rho_{k, N}\right) \leq 1 \\
    & \rho_{k,N}\succeq 0.
\end{aligned}
\end{equation}

Due to the finite accuracy of classical optimization algorithms, we also consider the dual problem of Eq.~\eqref{eq:sdp_pri_finite} to ensure the security,
\begin{equation}\label{eq:SDP_dual_finite}
\begin{aligned}
& \min_{\vec{\lambda} \succeq 0, \vec{\eta} \succeq 0, \xi \geq 0} -\sum_{j=1}^m \lambda_j p_j +\sum_{j=1}^m \eta_j p_j^L +\xi \\
\mathrm{s.t.} \quad & M_{k,N} +\sum_{j=1}^m (\lambda_j -\eta_j) M_{j,N} -(\xi+1) I \preceq 0, 
\end{aligned}
\end{equation}
where $(\vec{\lambda}, \vec{\eta}, \xi)$ is the dual variable and 
\begin{equation}
p_j^L = p_j -  \frac{\langle n \rangle }{N} || M_{j,\bar{N}}||_\infty.
\end{equation}

Suppose the optimal value of Eq.~\eqref{eq:SDP_dual_finite} is $d_N^*$ and that of Eq.~\eqref{eq:SDPprimalSI} is $p^*$, then $d_N^*\geq p^*$ by Eq.~\eqref{eq:upper_bound_finite}, which means the asymptotic randomness per sample has a lower bound 
\begin{equation}\label{eq:randomnessasym}
H_{\mathrm{min}}(A|E) \geq -\log_2 d_N^*.
\end{equation}





\section{Simulation}

In this section we make a simulation on the randomness lower bound by considering a practical phase-insensitive detector, for example, a time-multiplexed single photon detector (TMD) \cite{achilles2004photon}. We need to construct a theoretical model for the detector, i.e., to simulate the POVM elements $M_j$, which enables us to directly calculate the probabilities of measurement outcomes $p_j$. Considering that each $M_j$ is a diagonal matrix, we only need to simulate the diagonal term $\theta_j^{(n)}$. Notice that $\theta_j^{(n)} = \mathrm{tr}(\ket{n}\bra{n}\sum_{l=1}^\infty \theta_j^{(l)}\ket{l}\bra{l})$. This means that the diagonal term $\theta_j^{(n)}$ is the probability of obtaining the $j$-th outcome given an input Fock state $\ket{n}$. Then we simulate this conditional probability for a time-multiplexed detector.

\begin{figure}[hbt]
\centering
\includegraphics[width=0.5\textwidth]{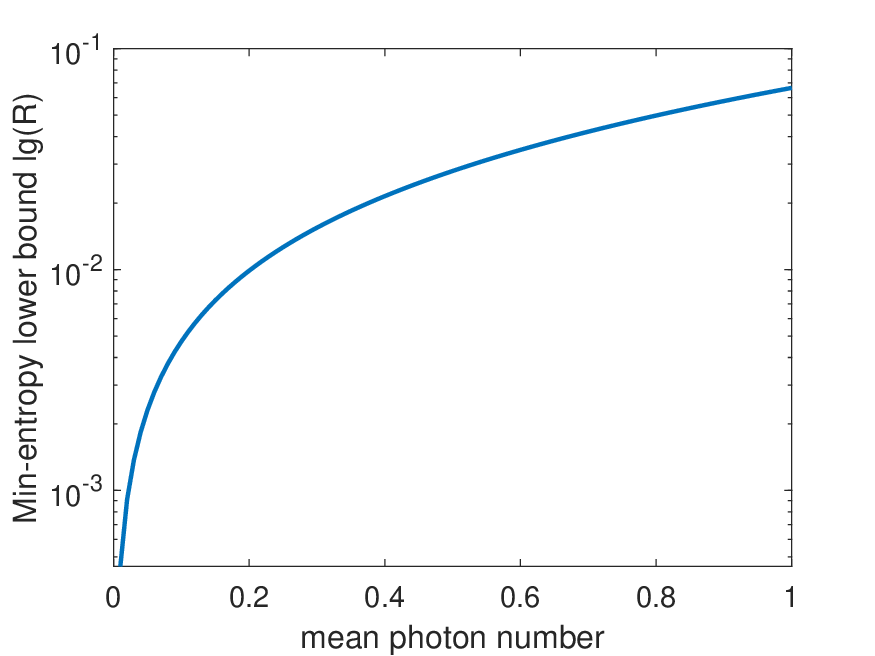}
\caption{Simulation of the randomness lower bound versus the mean photon number of the source.}
\label{fig:simulation}
\end{figure}

The time-multiplexed detector is designed to resolve multiple photon numbers through time-multiplexing techniques. It employs a series of single-mode optical fibers of varying lengths combined with symmetric $2 \times 2$ fiber couplers. When a light pulse enters the detector, it is split into multiple (usually the power of 2) temporal modes. This procedure can be viewed as a simple stochastic model where $n$ balls fall into $N_{\text{mode}}$ bins and exactly $j$ bins are occupied. If we assume each ball fall in each bin with equal probability. Then this situation corresponds to the Stirling number of the second kind in combinatorics, which characterizes the number of ways to partition a set of $n$ objects into $j$ non-empty subsets and is denoted by $S(n,j)$. The closed form of $S(n,j)$ is
\begin{equation}
S(n,j) = \sum_{i=0}^j \frac{(-1)^{j-i}i^n}{(j-i)!i!}.
\end{equation}
Then the probability of exact $j$ occupation is 
\begin{equation}
\theta_j^{(n)} = \frac{C_n^j j! S(n,j)}{N_{\text{mode}}^n},
\end{equation}
which generalizes the result in \cite{achilles2004photon}. In our simulation, we choose a coherent state $\ket{\alpha}$ as the source state, which is given by a superposition of Fock states,
\begin{equation}
\ket{\alpha} = e^{-\frac{|\alpha|^2}{2}}\sum_{n=0}^\infty \frac{\alpha^n}{\sqrt{n!}}\ket{n}.
\end{equation}
Then the probability of obtaining the $j$-th outcome is 
\begin{equation}
p_j = \mathrm{tr}(\ket{\alpha}\bra{\alpha} M_j).
\end{equation}
In the following, we calculate the randomness lower bound by choosing a set of practical parameters: the photon number cutoff $N=20$, number of POVM elements $m=10$, the number of temporal modes $N_{\text{mode}}=2^5$. We choose a weak coherent state source with mean photon number less than 1.
The simulation result is given in Fig.~\ref{fig:simulation}. It turns out the randomness lower bound can surpass $10^{-2}$ bit per sample for commonly used weak coherent state sources and TMDs.

\section{Conclusion}
We propose a simple CV-SI-QRNG scheme and developed a novel numerical security analysis framework for it. By eliminating the dimension assumption on the unknown quantum state, our framework extends the applicability of CV-SI-QRNG protocols. The key innovation lies in a dimension-reduction technique that maps infinite-dimensional quantum states and measurements to finite-dimensional SDP problems, ensuring computational feasibility without sacrificing security.

Our framework is compatible with a wide range of practical detectors, including phase-insensitive single-photon detectors, which simplifies implementation compared with conventional schemes. Simulations validate the robustness of the framework, showing its ability to achieve valid randomness lower bounds and high randomness generation rates. These advances address longstanding challenges in CV-SI-QRNG security analysis, providing a powerful tool for designing and analyzing future protocols. Our work thus represents a significant step toward the practical deployment of secure, high-performance SI-QRNGs in real-world applications.
\section*{Acknowledgement}
This work was supported in part by the National Natural Science Foundation of China Grants No. 61832003, 62272441, 12204489, and the Strategic Priority Research Program of Chinese Academy of Sciences Grant No. XDB28000000.

\appendix
\section{Derivation of the dual problem}
In this appendix, we derive the dual problem corresponding to the finite-dimensional SDP defined in Eq.~\eqref{eq:sdp_pri_finite} using the Lagrange duality framework.

The primal problem is given in Eq.~\eqref{eq:sdp_pri_finite}
To construct the Lagrange function, we introduce the dual variables: $\lambda_j \geq 0$ and $\eta_j \geq 0$ for the lower and upper bounds of the constraints on the measurement probabilities, respectively,
$\xi \geq 0 $ for the normalization constraint,
$Z_k \succeq 0$ for the semi-definite constraints on 
$\rho_{k,N}$
The Lagrange function is:
\begin{equation} \begin{aligned} L(\rho_{k,N}, \lambda_j, \eta_j, \xi, Z_k) &= \sum_{k=1}^m \mathrm{tr}(\rho_{k,N} M_{k,N}) + 1 - \sum_{k=1}^m \mathrm{tr}(\rho_{k,N})  \\ &\quad - \sum_{j=1}^m \lambda_j \left( \mathrm{tr}\left(M_{j,N} \sum_{k=1}^m \rho_{k,N}\right) - p_j\right) \\ &\quad - \sum_{j=1}^m \eta_j \left(p_j^L - \mathrm{tr}\left(M_{j,N} \sum_{k=1}^m \rho_{k,N}\right)\right) \\ &\quad - \xi \left(1 - \mathrm{tr}\left(\sum_{k=1}^m \rho_{k,N}\right)\right) \\ & \quad - \sum_{k=1}^m \mathrm{tr}(Z_k \rho_{k,N}). \end{aligned} \end{equation}
Reorganizing terms yields:
\begin{equation} \begin{aligned} & L(\rho_{k,N}, \lambda_j, \eta_j, \xi, Z_k) = - \sum_{j=1}^m \lambda_j p_j + \sum_{j=1}^m \eta_j p_j^L + \xi \\ & + \sum_{k=1}^m \mathrm{tr}\left[\rho_{k,N} \left(M_{k,N} + \sum_{j=1}^m (\lambda_j - \eta_j) M_{j,N} - (\xi + 1) I - Z_k\right)\right]. \end{aligned} \end{equation}

To ensure the Lagrange function is bounded from above, the coefficient of $\rho_{k,N}$ 
must satisfy the following semi-definite constraint:
\begin{equation} M_{k,N} + \sum_{j=1}^m (\lambda_j - \eta_j) M_{j,N} - (\xi + 1) I \preceq 0.
\end{equation}
Additionally, the dual variables must satisfy 
$\lambda_j \geq 0$, $\eta_j \geq 0$ and $\xi \geq 0 $.
The dual objective is obtained by minimizing the Lagrange function over the dual variables:
\begin{equation} \min_{\vec{\lambda} \succeq 0, \vec{\eta} \succeq 0, \xi \geq 0} - \sum_{j=1}^m \lambda_j p_j + \sum_{j=1}^m \eta_j p_j^L + \xi.
\end{equation}
Combining the objective and constraints, the dual problem is:
\begin{equation} \begin{aligned} & \min_{\vec{\lambda} \succeq 0, \vec{\eta} \succeq 0, \xi\geq 0} -\sum_{j=1}^m \lambda_j p_j + \sum_{j=1}^m \eta_j p_j^L + \xi \\ \mathrm{s.t.} \quad & M_{k,N} + \sum_{j=1}^m (\lambda_j - \eta_j) M_{j,N} - (\xi + 1) I \preceq 0. \end{aligned} \end{equation}
This completes the derivation of the dual problem in Eq.~\eqref{eq:SDP_dual_finite}.
\twocolumngrid
\bibliography{bibsemidiqrng}
\end{document}